\documentclass[conference]{IEEEtran}
\IEEEoverridecommandlockouts
\usepackage{cite}
\usepackage{amsmath,amssymb,amsfonts}
\usepackage{algorithmic}
\usepackage{graphicx}
\usepackage{textcomp}
\usepackage{xcolor}
\def\BibTeX{{\rm B\kern-.05em{\sc i\kern-.025em b}\kern-.08em
    T\kern-.1667em\lower.7ex\hbox{E}\kern-.125emX}}
\begin{document}

\title{A short review on qudit quantum machine learning\\
\thanks{This work was supported by the S\~ao Paulo Research Foundation (FAPESP),
Grants No. 2023/15739-3 and No. 2022/00209-6, the Coordination for the Improvement of Higher Education Personnel (CAPES), Grant No. 88887.829212/2023-00, the National Council for Scientific and Technological Development (CNPq), Grants No. 309862/2021-3, No.
409673/2022-6, and No. 421792/2022-1, and by the National Institute for the Science and Technology of Quantum Information (INCT-IQ), Grant No. 465469/2014-0.}
}

\author{\IEEEauthorblockN{Tiago de Souza Farias}
\IEEEauthorblockA{
\textit{Federal University of São Carlos}\\
\textit{Physics Department} \\
São Carlos-SP, Brazil \\
tiago.farias@ufscar.br}
\and
\IEEEauthorblockN{Lucas Friedrich}
\IEEEauthorblockA{\textit{Federal University of Santa Maria} \\
\textit{Physics Department}\\
Santa Maria-RS, Brazil \\
lucas.friedrich@acad.ufsm.br}
\and
\IEEEauthorblockN{Jonas Maziero}
\IEEEauthorblockA{\textit{Federal University of Santa Maria} \\
\textit{Physics Department}\\
Santa Maria-RS, Brazil\\
jonas.maziero@ufsm.br}
}


\maketitle

\begin{abstract}
As quantum devices scale toward practical machine learning applications, the binary qubit paradigm faces expressivity and resource efficiency limitations. Multi-level quantum systems, or qudits, offer a promising alternative by harnessing a larger Hilbert space, enabling richer data embeddings, more compact variational circuits, and support for multi-valued problem structures. In this work, we review the role of qudits in quantum machine learning techniques, mainly variational quantum algorithms and quantum neural networks. Drawing on recent experimental demonstrations, including high-level superconducting transmons, qutrit-based combinatorial optimization, and single-qudit classifiers, we highlight how qudit architectures can reduce circuit depth and parameter counts while maintaining competitive fidelity. We further assess the evolving software ecosystem, from specialized simulators and differentiable-programming libraries to extensions of mainstream frameworks. We also identify key challenges in control complexity, noise management, and tooling maturity.
\end{abstract}


\section{Introduction}

This paper reviews the use of qudits in Quantum Machine Learning (QML), focusing on variational quantum algorithms (VQAs) and quantum neural networks (QNNs). We outline the theoretical advantages of qudits over qubits for machine learning and survey recent theoretical and experimental progress in qudit-based QML models. We also review available simulation software and toolkits for qudit-based quantum computing. Finally, we evaluate the benefits and challenges of adopting qudits in QML and highlight open questions for future research. The goal is to inform physicists, computer scientists, and people working in interdisciplinary fields of the current landscape of high-dimensional quantum computing in machine learning and to illustrate how qudits expand the horizons of QML.

\section{Qudits vs Qubits: Theoretical Advantages}

A qudit is a quantum system with $d$ orthogonal basis states $(|0\rangle, |1\rangle, ..., |d-1\rangle)$, generalizing the qubit ($d=2$) \cite{balantekin_properties_2024, nikolaeva_efficient_2024, wang_qudits_2020}. An $n$-qudit register thus spans a Hilbert space of dimension $d^n$, which grows faster in $n$ than the $2^n$ of qubits. This higher-dimensional state space yields a greater information capacity, for example, a single qutrit ($d=3$) can encode $\log_2 3 \approx 1.53$ bits of information, compared to exactly one bit for a qubit, and in principle allows QML algorithms to represent more data with fewer physical systems \cite{9430718,ACAR2025129404,valtinos2023,mandilara_classification_2024}. In practice, this means that many tasks which require multiple qubits (and hence deeper, more entangled circuits) can be carried out with shallower, simpler qudit circuits, reducing both the total number of entangling gates and the cumulative decoherence suffered on near-term devices \cite{preskill_quantum_2018}.

Beyond this compression, qudits offer a richer set of operations for quantum feature encoding. The algebra of $SU(d)$ provides additional generators (e.g., the eight Gell-Mann matrices for qutrits) that can be used to construct high-dimensional rotations and feature maps in a ($d^2-1$)-dimensional Bloch hypersphere \cite{pudda2024, farias2024}. For instance, the Gell-Mann feature map embeds classical inputs into an 8-dimensional space on a qutrit, enabling classifiers to capture subtle patterns with fewer parameters, much like increasing the width of a hidden layer in a neural network \cite{valtinos2023}. This enhanced expressivity has been demonstrated empirically: single-qudit classifiers can implement non-linear decision boundaries without entangling gates, and qutrit-based QNNs often reach a target accuracy with significantly fewer trainable parameters than their qubit counterparts.

Qudits also simplify the mapping of naturally multi-valued problems onto quantum hardware. Tasks such as three-class classification, d-ary combinatorial optimization, or the direct simulation of spin-1 models avoid the overhead of binary encodings and ancillary qubits, preserving problem structure and symmetry in the circuit itself \cite{mandilara_classification_2024, Vargas_Calder_n_2021, Jansen_2024, quditspin}. This native alignment can further reduce circuit depth and improve learning performance, since the variational ansatz needs fewer layers to explore the relevant subspace.

Finally, while qudit gates may exhibit higher individual error rates, due to more complex control and leakage channels, many quantum error-correction codes for qudits require fewer physical levels per logical unit, and specific protocols have shown increased noise tolerance using high-dimensional carriers \cite{PRXQuantum.4.030327, chizzini_quantum_2022}. Moreover, because qudit circuits can be much shorter, the net decoherence over an algorithm can decrease despite noisier gates \cite{jankovic_noisy_2024, srivastav_quick_2022, hartmann_nonlinearity_2025, dutta_noise-adapted_2025, krishna_towards_2019}. Thus, qudits offer a compelling trade-off: encoding more information per element and exploiting SU(d) embeddings can boost expressivity and compress circuits, making them a powerful tool for NISQ-era quantum machine learning \cite{hybrid1}.

Table \ref{tab:table1} summarizes key differences between qubit-based and qudit-based QML from a theoretical standpoint. Notably, qudits can reduce the elements and gates required for specific computations. A clear example is the multi-controlled gate: qudits can implement a Toffoli with fewer interactions than qubits \cite{toffoli_qudit, PhysRevA.75.022313}. An experimental study on IBM hardware found an order-preserving Toffoli decomposition with qutrits required only 4 two transmon operations (vs. 8 with qubits), achieving a modest fidelity improvement and faster execution \cite{galda2021}. Similarly, algorithms with inherently trinary or higher-arity variables (e.g., a 3-class classification or graph 3-coloring problem) can be mapped more naturally to qutrits, avoiding the overhead of encoding these into multiple qubits or unused basis states \cite{10313913}. This resource efficiency can translate to shorter circuit depth and fewer parameters to optimize, which is advantageous for noisy intermediate-scale devices.

\begin{table}[htbp]
\caption{Comparative Characteristics between qubits and qudits.}
\begin{center}
\begin{tabular}{|c|p{0.275\linewidth}|p{0.275\linewidth}|}
\hline
\textbf{Aspect} & \textbf{Qubits} & \textbf{Qudits} \\
\hline
\hline
\textbf{State space} & 2 levels & $d$ levels\\
\hline
\textbf{Information encoding} & 1 bit per qubit & $\log_2(d)$ bits per qudit\\
\hline
\textbf{Expressivity} & Requires more qubits or layers to represent complex functions. & Potentially represent complex patterns with fewer units or circuit layers.\\
\hline
\textbf{Circuit complexity} & Some operations need multiple qubits. Gate decompositions may require many elementary gates. & Certain multi-qubit operations can be done within one qudit or with fewer gates. Problems with natural base-$d$ variables can be implemented directly, reducing overhead.\\
\hline
\textbf{Hardware overhead} & More qubits needed to achieve a given Hilbert space dimension. & Fewer qudits can achieve the same computational space. This can mean fewer physical components for the same task.\\
\hline
\textbf{Noise and errors} & Well-characterized two-level errors; mature control for qubits in many platforms. & Tends to have higher error rates per unit. More complex error modes due to more levels.\\
\hline
\textbf{Software support} & Most QML frameworks and compilers assume qubits. & New libraries (QuForge, QuDiet) are being developed, but ecosystem is less mature.\\
\hline
\textbf{Error correction} & Established qubit QEC codes (e.g. surface code). & Qudit QEC codes exist in theory and may offer advantages but not yet implemented widely.\\
\hline
\end{tabular}
\label{tab:table1}
\end{center}
\end{table}

However, theoretical advantages come with trade-offs. Higher-dimensional quantum operations are generally more complex to control and may have lower fidelities. While qudits provide greater control of the Hilbert space, in practice, producing and manipulating qudits is often more difficult than qubits. The larger Hilbert space can also lead to more complicated error channels (e.g., leakage from computational subspace). Nonetheless, a growing body of literature suggests that in the QML context, where expressive, compact circuits are desired, the benefits of qudits can outweigh the costs for specific tasks \cite{qudit_advantage, lysaght_quantum_2024, de_oliveira_unconditional_2025}.

\section{Applications on quantum machine learning algorithms}

Variational Quantum Algorithms are a class of hybrid quantum-classical algorithms where a parameterized quantum circuit is optimized, via a classical optimizer, to extremize a cost function \cite{cerezo_variational_2021, qi_variational_2024}. Examples include the Variational Quantum Eigensolver (VQE) for finding ground states of molecules \cite{cerezo_vqe, fedorov_vqe_2022, peruzzo_variational_2014}, and the Quantum Approximate Optimization Algorithm (QAOA) for combinatorial optimization \cite{farhi2014, BLEKOS20241}. VQAs are central to QML, often serving as quantum analogues of neural network training with tunable circuit parameters. Incorporating qudits into VQAs can broaden the capability of the algorithm by exploiting higher-dimensional quantum search spaces and reducing the number of required qubits or gates \cite{Cao2024Emulating, roca-jerat_qudit_2024}.

One of the first implementations of qudit-based VQAs was to use a single four-level superconducting transmon to emulate two qubits in a chemistry problem \cite{Cao2024Emulating}. The authors implemented a VQE on a transmon qudit, with dimension 4, to simulate a two-qubit system. They compiled a two-qubit molecular Hamiltonian problem into qudit operations and ran the optimization on hardware, applying error mitigation for readout and decay errors. Remarkably, the final energy achieved was within chemical accuracy of the true ground state, demonstrating that a 4-level qudit can effectively stand in for two qubits in a VQE. This result validates that qudits are a practical alternative to qubits for variational algorithms on current devices. It also hints at a hardware efficiency: using one physical qudit instead of two qubits reduces crosstalk and gates between separate qubits, at the cost of more complex single-qudit control.

Another use case for qudits in VQE is to simulate systems that naturally live in a higher-dimensional state space. For instance, a spin-$1$ particle has three basis states $(m_s = -1, 0, +1)$, which maps directly to a qutrit \cite{quditspin}. Using qutrits to simulate spin-$1$ Hamiltonians avoids encoding spin states into multiple qubits. Similarly, vibrational modes or bosonic oscillators truncated to $d$ levels can be treated as qudits \cite{tacchino_proposal_2021}. By matching the simulation basis to the physical basis, one can simplify the variational circuit. While specific QML examples of this are still emerging, conceptually, a variational ansatz on qudits can more natively represent specific problems than qubit mappings.

QAOA benefits significantly when the problem has a natural formulation in base $d$ \cite{PhysRevA.107.062410}. A striking example is the graph 3-coloring problem, which assigns one of three colors to each node of a graph such that adjacent nodes differ in color. With qubits, solving this requires encoding the three colors into two qubits per node (since $2^2=4$ possibilities, we have one wasted state) or using binary encodings with penalty terms for the invalid color. In contrast, with qutrits, one can assign each node to a qutrit (states 0,1,2 corresponding to colors) and implement the coloring constraints directly. 

A recent study \cite{10313913} formulated a qutrit-based QAOA for graph 3-coloring and compared it to an optimized qubit encoding. The qutrit QAOA achieved the same or better solution quality with shallower circuits and half the number of two-qubit gates per layer. In numerical simulations on random graphs, the qutrit version consistently had a higher probability of sampling a correct coloring than the qubit version for the same number of QAOA layers. Even more, the qutrit approach found better solutions using fewer layers, indicating it could reach a given performance level with less circuit depth. This points to an expressive advantage: the qutrit-parametrized circuit can explore the solution space more effectively per layer of the ansatz. The authors note that fewer layers and gates also mitigate some noise impact, partially offsetting multilevel gates generally higher error rates.

Beyond coloring, any optimization problem with a natural $d$-ary variable domain qualifies for qudit QAOA. Early theoretical work \cite{smith_programming_2022} showed asymptotic gate count improvements for general circuits using qutrits, which translates to QAOA and fewer gates to implement the same multi-controlled phase operations. 

It should be noted that optimizing VQAs with qudits might require different parameter-update rules. For example, the parameter-shift method for gradients \cite{PhysRevA.98.032309}, common in qubit VQAs, becomes more involved for higher-dimensional gates, sometimes requiring additional measurements per parameter \cite{wierichs_general_2022}. This is an active area of research. Nonetheless, the evidence, including hardware experiments and simulations, indicates that qudits can confer a meaningful advantage in variational QML by cutting down the resources needed and aligning quantum representations more closely with problem structure.

\section{Qudit-Based Quantum Neural Networks}

Quantum neural networks \cite{9137960} refer to parametrized quantum circuits or algorithms inspired by neural network architectures, used for tasks like classification, regression, or generative modeling. In many QNNs schemes, data is encoded into quantum states (a feature map step), then a layered parametric circuit is applied, and measurements yield the output, which is used to train the parameters. Replacing qubits with qudits in QNNs could enhance their modeling power and reduce the size of the quantum circuit needed for a given task.

One prominent line of work in this area is qudit-based classifiers. For instance, one recent work proposed a QNN-like model using a single qudit to perform binary classification via geometric rotations on the qudit’s state space \cite{mandilara2024}. In their approach, classical data points are mapped onto the surface of a Bloch hypersphere by a data-encoding rotation. Then, a trainable rotation is applied, and a projective measurement is made to decide the class. During training, the encoding and rotation parameters are adjusted via gradient descent to minimize a cost function (based on measurement expectation). Remarkably, they demonstrate that even a single qudit ($d=3$ or $4$) can solve nonlinear classification tasks with only a few adjustable parameters without entangling gates. This result highlights the significant expressive power of a lone high-dimensional quantum node, functioning as a multi-dimensional perceptron, where its rich geometry enables decision boundaries that would otherwise demand entangling multiple qubits.

Other researchers have explored quantum neural networks built from qutrit-based circuits with multiple layers. For instance, one work \cite{ACAR2025129404} introduced a QNN architecture composed of stacked parameterized qutrit gates, using qutrit-specific feature maps and rotations analogous to artificial neurons, loped an 8-dimensional Gell-Mann feature map for a qutrit and used it in a variational circuit classifier, showing improved classification accuracy on specific datasets compared to qubit-based encoding. The QNN leveraged both the higher state space and unique transformations available in $SU(3)$. For example, a qutrit rotation can entangle amplitude and phase across three basis states in ways a sequence of qubit rotations cannot easily replicate. This can be seen as quantum feature enrichment, potentially giving the QNN a head start in capturing data structure.

Whether these enhancements translate to concrete performance gains and resource savings is a key question. Comparative studies are beginning to answer that. For example, researchers systematically compared qubit vs qutrit variational QNNs on classification tasks, keeping the model architecture as equivalent as possible \cite{ACAR2025129404}. They examined different data encoding strategies (angle encoding, where features modulate rotation angles, vs amplitude encoding, where features set the amplitudes of basis states) in both $2$-level and $3$-level implementations. The findings indicated that qutrit-based QNNs can achieve equal or higher accuracy with fewer parameters than qubit QNNs for certain encodings. In particular, using simple angle encoding, the qutrit system reached a given accuracy with significantly fewer circuit layers/parameters, highlighting a more efficient use of parameter space. 

This supports the hypothesis that a single qutrit carries more expressive power than a single qubit, making the network more compact. Their results also suggested an advantage for qutrits for amplitude encoding, though the difference was less pronounced. Overall, this comparative study provides quantitative evidence that higher-dimensional quantum units improve model efficiency in QML. It also serves as a benchmark for future designs, indicating where qudits yield the most significant gains, such as encoding schemes, dataset types, etc.


Additionally, a qudit can implement data re-uploading (re-encoding inputs in multiple layers) more efficiently \cite{wach_data_2023}. In a data re-uploading scheme, one repeatedly encodes features and applies trainable gates in alternating layers. A recent work \cite{qutrit_transmon} used a transmon qutrit to realize a data re-uploading classifier with an encoding optimization technique. They tried various ways of encoding classical inputs into a qutrit and trained the circuit parameters and the encoding strategy to maximize accuracy. The optimized qutrit QNN achieved high accuracy on a ternary classification task ($3$ classes), using significantly fewer circuit elements than an equivalent qubit circuit. Moreover, they successfully ran this optimized qutrit QNN on a noisy superconducting qutrit, demonstrating robust performance on hardware.

\section{Qudit Simulation and Software Tools}

Robust simulation software and programming frameworks are essential for designing and testing qudit-based QML algorithms. In recent years, the quantum software ecosystem has begun extending support to qudits, although it lags behind the extensive tools for qubit circuits. Here, we survey the available and emerging software platforms for simulating and programming qudits.

\begin{itemize}

     \item \textbf{Circ \cite{developers_cirq_2024}:}  Cirq is a Python quantum circuit framework natively supports qudits. In Cirq, a qudit is treated similarly to a qubit but with a specified dimension property. One can create a vector state and define custom gates that act on $d$-level systems. Cirq ensures that gates are only applied to qudits of matching dimension. Many standard gates are provided for prime dimensions, and users can build others.

    \item \textbf{MQT Qudits \cite{mato2024}:} an open-source extension of the Munich Quantum Toolkit that enables the design, simulation, and compilation of mixed-dimensional quantum circuits using qudits. It introduces a Python API for specifying circuits on heterogeneous qudit registers, offers simulation via both decision-diagram and tensor-network backends, and provides a modular, pass-based compiler that maps high-level algorithms to hardware-native gate sets, thereby leveraging the increased information density and richer gate repertoire of qudits to produce shorter, more expressive circuits while abstracting away platform-specific details.

    \item \textbf{QuDiet \cite{Chatterjee_2023}:} QuDiet is another recently introduced simulator focusing on hybrid qubit-qudit systems. It provides a way to simulate circuits that include qubits and qudits in the same register, reflecting that one might use qudits in some parts of an algorithm and qubits in others. QuDiet’s emphasis is on being user-friendly for higher-dimensional simulation.
    
    \item \textbf{QuForge \cite{farias2024}:} QuForge is a specialized library introduced in 2024 explicitly for qudit circuit simulation. It provides a user-friendly way to build quantum circuits with arbitrary qudit dimensions. QuForge includes a library of $d$-dimensional quantum gates and allows users to specify an overall circuit of qudits of chosen dimensions. Notably, QuForge is built on differentiable programming frameworks and supports GPU/TPU acceleration. This means it can handle larger state spaces efficiently and even compute gradients of circuits using automatic differentiation. It also implements sparse matrix methods to save memory. By constructing circuits as differentiable computational graphs, QuForge directly targets QML applications that need to optimize parameters, as it can interface with machine learning libraries for hybrid quantum-classical optimization.

    \item \textbf{QuTiP \cite{JOHANSSON20121760}:}  QuTiP is an open-source library primarily for simulating quantum dynamics, but it also has a module for quantum circuits and processors. QuTiP inherently works with matrices and operators of arbitrary dimension, so it is well-suited to simulate qudits by constructing the appropriate Hamiltonians or gate operators. While it may not have a high-level circuit API for qudits, users can manually build unitary matrices for multi-level gates and propagate states.

    \item \textbf{Other high-level frameworks:} On the algorithm development front, libraries like Pennylane \cite{bergholm2022} or Qiskit \cite{javadiabhari2024} do not officially support qudits in their core API. However, one can often integrate lower-level control. In Qiskit, advanced users have used Qiskit Pulse to create qutrit gates on IBM hardware. One could use its unitary simulator for simulation by manually constructing a 3-level gate as a 3x3 matrix acting on a single qubit’s 3-level subspace. Pennylane can interface with Cirq, so a Pennylane user can define a Cirq device that has qudits and then use Pennylane’s high-level optimization with it, albeit with some limitations.
\end{itemize}

The landscape of software is growing, but standardization is still lacking. One of the efforts pushing this is developing a common quantum assembly language that supports qudits. OpenQASM 3.0 \cite{Cross_2022}, for instance, has types that could be extended to non-binary quantum registers, but implementations are nascent. The research community is also exploring new simulation techniques to handle larger quantum circuits efficiently. This is crucial for studying qudit QML at scale, since a straightforward simulation of many qudits becomes intractable quickly (the state space grows as $d^n$). Techniques that exploit structure (like tensor networks or symmetries in qudit circuits) may help push the boundary of classically simulable qudit models.

\section{Challenges and open questions}

Controlling and scaling multi-level quantum systems introduces a host of interrelated hardware challenges. As the number of accessible energy levels increases, so do the possible transitions and cross-talk pathways, making the calibration of multi-level gates both labor-intensive and reliant on sophisticated pulse-shaping techniques. 

Although recent hardware demonstrations have achieved respectable fidelities through bespoke control schemes, extending these protocols to tens of qudits, and maintaining them over extended runtimes, will demand new automated tuning algorithms and perhaps real-time feed-forward control. Compounding these control hurdles are the intrinsic noise and decoherence characteristics of higher energy levels. In superconducting transmons, for example, the relaxation time for the $|2\rangle \rightarrow |1\rangle$ transition is typically shorter than for $|1\rangle \rightarrow |0\rangle$, and leakage into non-computational states further degrades gate performance. While offering longer-lived multi-level states, trapped-ion platforms must contend with off-resonant coupling between closely spaced levels that can introduce spurious errors. As a result, matching the per-gate fidelities of qubits remains a formidable challenge: fidelity gaps that are tolerable for small circuits quickly become prohibitive as qudit-based algorithms grow in size.

Parallel to these hardware constraints, the software ecosystem for qudits still lags behind the mature tooling available for qubits. Most quantum developers think in binary logic, so adopting high-dimensional systems requires new abstractions, libraries, and workflows. With few standardized frameworks or off-the-shelf circuit templates for multi-level operations, newcomers face a steep learning curve, and established QML techniques, from variational ansatze to error-mitigation strategies, must be re-examined in the qudit context. This combination of hardware complexity and limited software support continues to slow the integration of high-dimensional computing into mainstream quantum research.

On the theory side, we still lack a comprehensive understanding of when and why a qudit-based QML model outperforms a qubit one. Current evidence is often case-by-case. It remains an open question: Is there a generalizable quantum advantage to using qudits for machine learning? For example, can we prove that a qudit QNN can approximate certain functions 
more efficiently than qubit QNNs? Or are there fundamental trade-offs that cap the advantage? Developing theoretical measures of model capacity (like effective dimension, entanglement capability, etc.) for qudit circuits vs qubit circuits is needed to answer these questions. Additionally, more work is required to understand the training landscapes of qudit circuits. As demonstrated in \cite{friedrich_barren_2025}, increasing the dimension of qudits exacerbates the BP problem. This raises the question: can existing BP mitigation techniques be applied to models based on qudits, or would developing new, specialized approaches be necessary?

\section{Conclusion}

Qudits are emerging as a promising avenue to enhance quantum machine learning models. This review has highlighted how qudits can offer theoretical advantages by expanding the Hilbert space per physical system and enabling more compact or expressive variational circuits. In variational algorithms like VQE and QAOA, using qutrits and other qudits has reduced circuit depth and parameter counts while tackling problems in their natural state space. In quantum neural networks and classifiers, qudits have demonstrated the ability to learn complex patterns with fewer units and even achieve tasks unattainable by equivalent qubit networks. Recent experimental research have moved qudits from theory to practice, affirming that the high-dimensional approach to QML is feasible on today’s quantum devices.

We have also discussed the current limitations and challenges. Control and error rates for qudits are not yet on par with qubits, and significant engineering and calibration effort is required to harness multi-level systems at scale. While improving with tools like Cirq and QuForge, the software ecosystem still trails the qubit world in maturity. These challenges, however, are being actively addressed by the community, motivated by the clear potential rewards. Developing efficient qudit simulation techniques and error mitigation strategies will be critical to further progress.

 As research continues, we anticipate more hybrid approaches (mixing qubits and qudits), more specialized algorithms tailored to qudit strengths, and eventually, larger-scale demonstrations. Continued interdisciplinary efforts could lead to QML algorithms that learn faster, require fewer quantum resources, and solve problems more naturally than their qubit-only counterparts, unlocking higher-dimensional quantum intelligence for complex computational tasks.

\bibliographystyle{IEEEtran}
\bibliography{IEEEabrv, bibliography}

\end{document}